# Scavengers in the human-dominated landscape – an experimental study

Sourabh Biswas[1], Tathagata Bhowmik[2], Kalyan Ghosh[3], Anamitra Roy[1], Aesha Lahiri[1],

Sampita Sarkar[1] and Anindita Bhadra[1*]

**Affiliation**:

[1] Department of Biological Sciences, Indian Institute of Science Education and Research Kolkata, Nadia, West Bengal, India.

[2] Department of Zoology, West Bengal State University

[3] Department of Zoology, University of Burdwan

[*]**Address for Correspondence**

Behaviour and Ecology Lab, Department of Biological Sciences,

Indian Institute of Science Education and Research Kolkata

Mohanpur Campus, Mohanpur, Nadia

PIN 741246, West Bengal, India.

Phone no: +91 33 6136 0000 ext 1223

[*]Corresponding author

E-mail: abhadra@iiserkol.ac.in

## Abstract

Rapid urbanization is a major cause of habitat and biodiversity loss and human-animal conflict. While urbanization is inevitable, we need to develop a good understanding of the urban ecosystem and the urban-adapted species in order to ensure sustainable cities for our future. Scavengers play a major role in urban ecosystems, and often, urban adaptation involves a shift towards scavenging behaviour in wild animals. We carried out an experiment at different sites in the state of West Bengal, India, to identify the scavenging guild within urban habitats, in response to human provided food. Our study revealed a total of 17 different vertebrate species were identified across sites over 498 sessions of observations. We carried out network analysis to understand the dynamics of the system, and found that the free-ranging dog and common mynah were key species within the scavenging networks. This study revealed the complexity of scavenging networks within human-dominated habitats.

**Introduction:**

A recurrent and persistent theme of discussion across disciplines and in public fora today is climate change and its impact on human lives as well as the ecosystem. While a wide spectrum of research focuses on mitigation of climate change, disaster risk management and conservation of natural ecosystems, relatively less research focuses on the urban ecosystem as a habitat for biodiversity conservation. Rapid urbanization poses a threat to most wild species, but some species are nevertheless able to cope with the changing environment and colonize the new niches available in the urban ecosystem to thrive. While rapid urbanization is a very recent phenomenon in the history of the earth, alterations to the environment by humans is perhaps as old as human history. Anthropogenic food subsidies became available to animals around the hunter-gatherer communities of early humans. In the Neolithic era, with the advent of agriculture and domestication of wild animals, human settlements became a source of additional nutrition for various commensal species (Agudoet et al., 2010; Chamberlin et al., 2005). The explosion of the human population led to major alterations in habitats, through deforestation, agricultural practices and production of a large amount of waste, which acted as food subsidies for various species of animals (Oro et al., 2013).

Anthropogenic food subsidies, like refuse from food production, garbage dumps, fisheries discards, crop leftovers, feeding restaurants for scavengers, feeding stations for animals, etc. are predictable resources (Oro et al., 2013). Due to this predictable nature of the resources, they have been exploited widely by a large number of urban-adapted species, which typically demonstrate phenotypic and behavioural plasticity, variable niche adaptability and dispersal strategies (Luna et al., 2021). This temporal and spatial predictability of the food subsidies impact the increased survival, population breeding site availability, population growth and body mass of urban scavengers (Oro et al., 2013). Scavengers are animals that feed on dead and decaying matter. In the wild habitats, they typically feed on carcasses. However, in the urban ecosystem, scavengers are generalist feeders that feed on any available food, typically from anthropogenic sources. Waste generation in and around human settlements has increased substantially in the last few decades, and the volume of waste is predicted to double within the next twenty years (Hoornweg & Bhada-Tata 2012). Around the world, over 3 million tons of waste are generated regularly, which is expected to reach the staggering amount of 6 million tons per day by 2025 (Hoornweg et al., 2013). Needless to say, this is expected to impact the population dynamics and behaviour of urban scavengers and the ecology of wild species substantially.

A few studies have focused on the carrion-feeding (Corté S-Avizanda et al., 2012) species associated with scavenging communities in habitats with different predictable anthropogenic food subsidies (Oro et al., 2013). Some studies have focused on the birds associated with landfills and large garbage dumps. Very little information is available about scavenging communities in urban habitats in South Asia. Due to the nature of human lifestyle in the present times, the urbanization gradient across habitats is no longer very steep, and clear demarcations of urban and rural habitats no longer exist. Hence, it is better to consider landscapes in terms of levels of anthropogenic usage, rather than a simple dichotomy of rural and urban, based on lifestyles. Katlam & Prasad (2018) have investigated the scavengers in the forest fringe areas in Uttarakhand associated with seasonal tourist influx. All of the studies have focused on predictable food subsidies. However, scavenging response to unpredictable food subsidies in human-dominated habitats is yet to be explored.

The structure of the community is an important factor for ecosystem functioning and an essential driver for the stability of the community and its conservation (González et al., 2020). Vertebrate scavengers in terrestrial habitats play a crucial role in ecosystem functioning as they help accelerate decomposition and nutrient cycling, stabilize food webs, help to control disease transmission and pest expansion (Beasley et al., 2019; Inger et al., 2016). Several studies have described the community structure through network analysis (Chakraborty et al., 2021; González et al., 2020; Sebastián-González et al., 2016) which is an effective tool to summarize the community assemblage pattern numerically and compare communities statistically (Bascompte & Jordano, 2007). Community structures are analyzed in nested patterns, using specialization and centrality levels for understanding of the importance of different species in the ecological community (Bascompte & Jordano, 2007; Sebastián González et al., 2020). Understanding the structures of scavenging communities can be crucial in ecosystem management and conservation, and can be an important factor contributing to the policy level decisions to ensure sustainable urban management.

This study aimed to document the diversity of the scavenging guild associated with daily human-generated waste by providing vegetarian and non-vegetarian food typical to Indian households ad-hoc at various sites. This helped us to document the scavenging guild and also study the response pattern of different species to the food provided. Finally, we analyzed the network structure of scavengers in response to human generated food source and identified the keystone scavenger species in the human-dominated habitat. Since free-ranging dogs (*Canis lupus familiaris*) are an integral part of all human habitats in India, and have been adapted to living among humans for perhaps the longest time, we tried to specifically understand the role of the free-ranging dogs within this scavenging guild, and how they impact the diversity of scavengers exploiting the ad-hoc resources created.

## Methods

### Study area

The experiment was conducted between 2020 and 2021 in different parts of West Bengal, India. Experimenters choose sites randomly in and around the human habitation and away from existing resources. The study sites were distributed in different parts of West Bengal, India (Table S1, Figure S1).

### The scavenging guild experiment

The specific sites for the experiment were selected away from the existing regular resource sites that were seen to be used for scavenging by free-ranging dogs, through reconnaissance surveys. The experimenter upturned a bowl of boiled rice or roti pieces mixed with either vegetarian curry (VF) or non-vegetarian (egg, fish or chicken) curry (NVF). This is the most common food of people across India, and home cooked food was used for all the experiments. The bowls used were typically of a volume of approximately 400 ml. Video recording was initiated before placing the food from a distance of 5-10 meters, and recording was continued for one hour or until all the food provided was consumed by scavengers. The food type (VF/NVF) for the first trial was picked randomly for each site, and once a food type was chosen, the same was provided twice a day during morning and afternoon sessions for ten consecutive days, followed by the other food type over the next ten days.

**Data collection**

The videos were decoded to extract the following information:

(i) Identity of scavengers: Any individual that was seen to approach and sniff and/or eat the food was designated as a scavenger. The species identities of the scavengers were noted.

(ii) Latency: The time taken for an individual to approach and interact with the food, from the time of dropping the food, was considered as latency (in seconds).

Data on temperature, wind speed, cloud cover and rain corresponding to the experimental sessions were collected from www.worldweather.com.

**Statistical analysis**

(i) Non-parametric tests: R statistical (R Studio) software was used for all the statistical analyses. Shannon diversity index (H′) of the scavenger community in each session was calculated using the *vegan* package (Oksanen et al., 2007). Mann-Whitney U test was used to compare scavenger diversity and latency of the first responders (FR) across the sessions (Morning/Afternoon) and food type (VF/NVF). Diversity and latency of the FR were compared across sites and scavengers as FR using the Kruskal-Wallis test. Post-hoc Dunn's test was performed to check for any significant difference in both latencies of FR and session diversity across sites and scavengers.

(ii) Modeling: A GLMM (Generalized linear mixed model) analysis was conducted to understand the effect of the first responder (species identity) and temperature on the latency of the FR using normal distribution after log-transforming the data for latency of the first responder using "lme4" (Bates et al., 2015) package of R. Session diversity of the species was separated into three categories; low (0 – < 0.5 H′), medium (0.5 – <1 H′), high (≥1 H′). Ordinal mixed logistic regression (OMLR) was carried out to understand (a) the effect of the first responder species, sessions, and temperature on the session diversity category using "ordinal" package (Christensen RHB, 2019); (b) the effect of dogs as responders and sessions on the session diversity category and (c) the effect of common mynas as FR and session on the session diversity category. Experimental sites were considered as the random effect. The null and full models were compared for all the models. Dispersion of the models and residual diagnostics were checked using the "performance" package of R (Lüdecke et al., 2021). The alpha level was kept at 0.05 throughout the analysis.

(iii) Network analysis: Sites and scavenger interaction networks were built using the "bipartite" package in R (Dormann et al., 2008). Scavengers and sites were placed in columns and rows, respectively, to construct a quantitative interaction matrix. Nestedness and network specialization were computed for the composite network (combined VF and NVF) and also separately for VF and NVF. A weighted matrix of scavenger response was used to calculate the 'nestedness metric based on the overlap and decreasing fill' (NODF) (Almeida-Neto et al., 2008). NODF was computed using 'nestednodf' command in the 'vegan' package (Oksanen et al., 2007) in R. NODF values range between '0' (no nestedness) to 100 (complete nestedness). Complementarity specialization ($H'_2$) was computed for weighted matrix (scavenger response) using the 'H2fun' command in the 'bipartite' package in R. The range of ($H'_2$) values fall between '0' (no specialization) and '1' (complete specialization).

A virtual exclusion method was executed to determine the significance of different scavengers and sites in the respective networks. Each scavenger species was eliminated from the data to compute the NODF and $H'_2$ values of the network, and then included back before eliminating the next species. Using this method, the significance of each scavenger species was evaluated in terms of network nestedness (NODF) and network specialization ($H'2$).

Three species level indices– normalized degree (ND), closeness centrality (CC) and betweenness centrality (BC) were computed for the weighted composite network and individually for VF and NVF networks. The degree of a species was divided by the number of sites and was used to normalize the three indices (González et al., 2010). A relationship between normalized degree, betweenness centrality and closeness centrality for both the scavengers and sites were calculated using 'Pearson' correlation in R.

**Results**

**Scavenging Guild**

The experiment was performed in 498 different sessions, out of which 250 were in the morning, and 248 were in the afternoon. A total of 240 sessions were completed with VF and 258 sessions with NVF. A total of 4403 individuals were recorded from the experimental sites. Species responses were significantly higher ($\chi^2 = 404.78$, df = 1, $p \leq 0.05$) for VF (2869), as compared to NVF (1534). A total of seventeen vertebrate species of scavengers were recorded in the study, out of which ten were avian species and six were mammals (Fig. S2, Fig. S3a). Fifteen species of scavengers were recorded in VF (Fig. S3b), while sixteen species were recorded in NVF (Fig. S3c). There was considerable overlap of species observed in the different sites and for food types. Overall, House crows (*Corvus splendens)*, sparrows (*Passer domesticus*) and common mynas (*Acridotheres tristis*) were the most common scavengers. The most abundant species responding to VF and NVF were sparrows, house crows, common mynas and house crows, dogs, common mynas respectively. Cows (*Bos taurus*) only responded to VF while white-breasted waterhen (*Amaurornis phoenicurus*) and domestic ducks (*Anas platyrhynchos domesticus*) only responded to NVF.

**Scavenger diversity of Sessions**

Species diversity of session varied across the sites (Kruskal-Wallis rank-sum test: $\chi^2 = 173.36$, df = 14, p = <0.001). Scavenger diversity was relatively higher at three sites of Bongaon area and the lowest at two sites of Durgapur (Fig. S4a).

Wilcoxon rank sum test revealed a significantly higher diversity of species (W= 27222.00, p = 0.01) in case of VF, as compared to NVF (Fig. S4b). The diversity of scavengers in the morning and afternoon sessions were comparable (W= 29888.00, p = 0.45).

When cat (*Felis catus*), common myna and jungle myna (*Acridotheres fuscus*) were the FR, a significant increase in the expected value of diversity on the log odds scale was observed, given that all the other variables in the model were held constant, compared to dogs as FR (Fig. S4c, Table. S2). This suggests that the dogs play a significant role in the scavenging community. We witnessed a decrease in the expected value of diversity on the log odds scale in case of temperature but the session didn't affect the expected value of diversity (Table. S2).

Dogs as no responder (NR) and second responder (SR) showed a significant increase in the expected value of diversity category on the log odds scale compared to dogs as FR (Fig. 1, Table. S3). The expected value of diversity category on the log odds scale positively increased in the afternoon session compared to morning (Table. S3). Common myna was the most ubiquitous scavenger that could be seen in all habitats. When common myna was the FR, we did not see any effect on the expected value of diversity on the log odds scale, when all the other variables in the model were held constant (Table. S4).

**Latency**

Kruskal-Wallis test revealed a significant difference in latency of the FR across sites ($\chi^2$ = 209.35, df = 14, p < 0.001). Latency was highest at three sites of Bongaon and lowest at two sites of Durgapur (Kruskal-Wallis rank-sum test: $\chi^2$ = 221.60, df = 14, p < 0.001) (Fig. S5a). Latency of the FR were comparable across the morning and afternoon sessions (Wilcoxon rank-sum test, W = 31188, p = 0.9). Latency of FR were also comparable across the veg and non veg food (Wilcoxon rank-sum test, W = 30302, p = 0.7) (Fig. S5b). While all species were included in the estimation of diversity, greater coucal, cow, jungle babbler, pied myna, goat and squirrel were not considered as FR in the comparison of latency using the GLMM and OMLR analysis that follow, as they contributed to ≤ 1% of the total sample size, and were considered as outliers. Latency of FR were significantly different across species (Kruskal-Wallis rank-sum test: $\chi^2 = 173.36$, df = 14, $p < 0.001$). Latency of dogs as FR was the lowest (2.32 ± 6.73s), whereas red vented bulbul showed the highest latency (11.41 ± 15.60s) (Fig. S5c).

GLMM analysis revealed that the variation in latency of the FR was evident across scavengers. Temperature positively affected the latency of the FR (Table. S5). The latency of red-vented bulbul, cat, common myna, crow and jungle myna as FR were significantly higher compared to the latency of dog as FR (Table. S5)

Female dogs exploited the resource more than male dogs ($\chi^2 = 37.453$, df = 1, $p \leq 0.0001$). The maximum number of species exploiting the resource was reached within 15 minutes of providing the food, over a period of five days (Fig. 2).

The observed site and scavenger composite network (Fig. 3) combining VF and NVF showed moderate nestedness (NODF= 55.84) and moderate specialization ($H'_2 = 0.53$). The network structure did not differ for network nestedness (NODF = 52.54 and NODF = 56.02 in VF (Fig. S6) and NVF (Fig. S7), respectively) and network specialization ($H'_2 = 0.56$ and $H'_2 = 0.48$, respectively). $H'_2$ increased and NODF decreased after the virtual exclusion of dog and common myna from the network (Fig. 3). $H'_2$ decreased after excluding house crow and sparrow as scavenger species, as compared to the composite network (Fig. 4). Virtual exclusion of cat, red vented bulbul, palm squirrel, goat, house crow, jungle crow, greater coucal and cow increased NODF. We found a similar trend in the virtual exclusion of species in both VF and NVF (see Fig. S8 for details).

70% of the scavenger species acted as a connector (BC > 0) in the combined network (Fig. S9b), whereas for the VF and NVF networks these values were 67% and 63% respectively (Fig. S9e & Fig. S9h) . This suggests that some connector species were unique to the two food type networks, thus contributing to the composite higher representation of connectors in the combined network. Dog and common myna share around 50% and 11% of the BC in the composite network; 46% and 30% of the BC in VF; 27% and 27% BC in the NVF networks, respectively (Fig. S9e & Fig. S9h). CC value of dog found to be the highest (0.068) among the other scavengers in the composite network (Fig. S9c). Dog and common myna shared the highest CC value (0.083 and 0.075) in both VF and NVF networks respectively (Fig. S9f & Fig. S9i). The relationship between ND & CC and ND & BC for the scavenger species was represented by a linear relationship ($R = 0.59$, $p = 0.01$ and $R = 0.81$, $p < 0.01$ respectively) (Fig.5).

## Discussion

The Anthropocene has been a period of large scale and far-reaching changes to the environment, some of which have led to the present-day climate crisis. We need to take cognizance of the fact that humans are constantly and irrevocably affecting the environment, leading to habitat loss, biodiversity loss and behavioural changes in many species of the wild, among others. The UN Sustainable Development Goals 2030, designed to achieve a sustainable future for our planet, cannot be realized without the combined attention of scientists, policymakers and members of the public. In the face of rapid urbanization, we need to work towards a sustainable solution that would enable us to retain and nurture biodiversity in the urban landscape. Understanding the ecology and behaviour of urban-adapted species can be way forward towards this goal.

In this study, we documented the scavenging community in human-dominated habitats, which were selected randomly, and had various levels of urbanization. While some sites were in areas of high human flux like markets, other were within residential neighbourhoods. The locations ranged from congested cities to relatively less congested townships. We documented a reasonably large number of species exploiting the newly created unpredictable resource, namely, human provided food. There was significant variation in the diversity and species composition arising due to the nature of the food provided – vegetarian or non-vegetarian, while the basic component of the food was carbohydrates. It was interesting to see that vegetarian food attracted more species than non-vegetarian food, which can most likely be attributed to a higher number of avian species exploiting the VF. House crow, common myna, sparrow and free-ranging dog were the most ubiquitous species, exploiting both vegetarian and non-vegetarian food, but of these, only dogs were present across all the sites. These species have been reported as scavengers from different parts of the world, and are also known for their adaptation to human-dominated habitats (Noreen & Sultan, 2021).

Free-ranging dogs had a significant impact on the scavenger diversity of the sessions. When dogs were the first to find the food source, the scavenger diversity of the session was lower compared to the sessions in which dogs were not the first responders. This is probably because dogs consumed most of the available food quickly, which did not leave much to be utilized by other species. Moreover, virtual exclusion of dogs resulted in an increase of the NODF value of the networks, underlining the key role that the dogs play in the scavenging network for both VF and NVF. Common mynah was found to be the next most important scavenger species, though they have not been reported in this role extensively yet. Mynahs are omnivorous birds, and the current result is suggestive of urban adaptation in these birds through opportunistic utilization of human generated resources.

Some species that were documented as scavengers in this study for the first time are jungle myna (*Acridotheres fuscus*), pied myna (*Gracupica contra*), spotted dove (*Spilopelia chinensis*), domestic goat (*Capra aegagrus hircus*), northern palm squirrel (Funambulus pennantii), greater coucal (*Centropus sinensis*) and duck (*Anas platyrhynchos domesticus*). Among these, the domestic goat and duck are accustomed to human food, due to their close association with humans as domesticates, but species like the spotted dove and greater coucal are neither domesticated, nor known to be scavengers. This observation thus raises questions regarding the impact of human food waste on the dietary habits of these birds, and needs further investigation.

Crow, common myna and sparrow comprised nearly 82% of the scavenging guild documented responding to the VF. These species are communal foragers, and both inter and intra-species co-feeding was observed during our experiments, which was probably the primary cause of their success in monopolizing the food provided.

Scavenging from human-generated waste made these species successful in the human-dominated landscape. Dogs are obligate scavengers due to their history of domestication (Axelsson et al., 2013). They depend mostly on human-generated waste for survival (Bhadra et al., 2016).

They are highly adapted to scavenging in the human-dominated landscape because of their high cognitive ability to find food and superior olfaction (Sarkar et al., 2019) and have become highly efficient scavengers in the human-dominated landscape. This might be the reason for discovering food sources better than other scavenger species across our study sites.

Three sites in the Bongaon area were more of a rural habitat, whereas two sites in the Durgapur area were highly urbanized. These sites in Bongaon had high latency, perhaps due to a lower density of scavengers, or due to higher fear of humans. This needs further investigation. The two sites in Durgapur had the fastest discovery time, probably because of a higher density of dogs and less fear of humans in these areas. A significant drop in the diversity of species in the presence of dogs as the first responder indicates the species' efficiency to monopolize human subsidized food. Cortez Avizanda et al., (2012) also reported a similar kind of result, where the diversity of the scavengers was relatively high before the arrival of the dominant specialist scavenger species Griffon Vulture (Gyps fulvus), at the food source.

The composite, VF and NVF networks revealed a moderate nestedness and moderate specialization compared to the highly nested network of scavenger assemblage in carrions from all over the world (Sebastián González et al., 2020; Selva & Fortuna, 2007). Like plant-pollinator mutualistic networks, the site-scavenger network can also be treated as a mutualistic association, in which a scavenger species exploits the site for a particular type of resource for their survival. In this case, the sites are not directly benefitted like the plants, but the reduction of wats at a site can be considered as an indirect advantage for the site. The presence of the scavengers maintain the ecosystem by increasing connectivity and stability in the food webs (Inger et al., 2016; Rooney et al., 2006; Wilson & Wolkovich, 2011), disbursement of nutrients in the ecosystem (Helfield & Naiman, 2001; Hewson, 1995; Schlacher et al., 2015) and providing direct sanitary benefit to humans by removing potentially dangerous reservoirs of bacteria (Ortiz & Smith, 1994; Vass A, 2001) and zoonotic pathogens fatal for humans and other animals (Monroe et al., 2015).

In the nested community, networks are reported to be asymmetric and cohesive in nature (Bascompte & Jordano, 2007; Sebastián González et al., 2020). Nestedness reflects the high-density interactions of core taxa. For example, in plant-animal mutualisms, generalist species form a dense core, and specialists are linked to it. In case of the scavenging guild, the highly urban-adapted species, or more efficient scavengers, are expected to be common across sites, while species with more limited scavenging capacity should occur in a subset of the sites (Selva & Fortuna, 2007; Sebastián González et al., 2020). The competition for resources is higher in the more nested networks of scavenger assemblages (Sebastián González et al., 2020). Such a scenario can occur in multiple cases: (i) in scarce resource availability (Selva & Fortuna 2007), (ii) the occurrence of specialized scavengers (e.g., vultures, Sebastián-González et al., 2016), (iii) presence of dominant scavengers (e.g., black bears, Allen et al., 2014) that can monopolize the resources or (iv) where the resources are relatively constant (Selva & Fortuna 2007; Sebastián González et al., 2020). An increase in competition and a decrease in biodiversity is observed in the networks with low nestedness (Bastolla et al., 2009; Basu P et al., 2016). Contribution to network nestedness can be assessed as the control capacity of a species in the mutualistic network (Cagua et al., 2019), and thus, network nestedness can be used to point out the key species that have an impact on the overall network integrity (Campbell et al., 2012; Cagua et al., 2019). The network becomes more stable with the decrease in network specialization, making the network more resilient (Bascompte & Jordano, 2007; Basu et al., 2021). Virtual exclusion of dogs and common myna from all the three kinds of networks (composite, VF and NVF networks) reduced the network's nestedness and increased network specialization. This suggests that the dog and common myna play important roles in the scavenging networks and removing these generalist species from the network might decrease the competition in the networks and make the network more vulnerable (Basu et al.,, 2021).

Species with high CC values impact the other species of the network, and species with a positive value of BC in the network are essential for the network's cohesiveness (Estrada, 2007; Martín González et al., 2010). The centrality indices identified the keystone species in the networks across our studied sites. The architecture and persistence are impacted by the species with a more central location in the network over the species at the edge (Cagua et al., 2019). Dogs and common myna contributed the maximum in centrality indices, thus establishing them as the most important species in the site-scavenger assemblage network. With a high percentage (70%, 67% and 63%) of connector species in the composite, VF and NVF networks reveal those species' importance in maintaining the site-scavenger network's integrity. Dog, common myna and house crow were found to be the essential species contributing to the stability of the VF network in the studied sites. However, at the same time, house crow and sparrow were considered specialist scavenger species in some of the selected studied sites in the VF and NVF site-scavenger network. In the composite network, dogs and common myna were found to be the generalist species in the site-scavenger network, contributing as the most important keystone species in the network. In contrast, other scavengers were found to be specialists.

In the composite network, normalized degree (ND) showed a strong linear correlation with both closeness and betweenness centrality. In both real-world and random networks, the various centrality measures are expected to be correlated (Li et al., 2015). However, a strong linear relationship between both ND and CC as well as ND and BC are not always observed in networks in nature (Martín González et al., 2010). All these are centrality measures, that indicate a node’s prominence in a network, and has the potential to play an important role in information transfer within the network. In our case, the more connected scavenger species are likely to influence the species diversity of the session.

Individual differences in response to anthropogenic food are influenced by age, sex or hierarchical ranking (Oro et al., 2013), which is perhaps reflected in our observation of female bias in dogs exploiting the food sources. Our findings indicate that female dogs are more responsive toward food. The alternative hypothesis can be that females need more energy than males. During pregnancy, females have many energy requirements. The study was performed during dogs' pre-mating (July-September) and post-mating (October-December) seasons. Female dogs might be more responsive to conserving energy for the gestation period. However, a third explanation could be a female biased population of dogs in the areas of study, which is unlikely, as an earlier study based on random sightings has reported no such bias in the population (Sen Majumder et al., 2014). We need further studies to arrive at a conclusion regarding this observation.

This study was a first attempt to document the scavenging community dependent on human generated and provided food in human-dominated landscapes. While we documented the commonly identified scavengers and were able to understand their roles in the scavenging network, we also identified several species to exploit the given food, which have never been documented as scavengers. We believe that this study builds a case for further investigation into the identification of scavenging species and their interactions within human-dominated habitats. Moreover, such studies can help us to understand how species might be undergoing dietary shifts under the pressure of urbanization, to adapt to the more urban landscape. Such knowledge will not only help to delve into the shifts in behaviour leading to urban adaptation, but will also shed light on the evolutionary question of how some species might have adapted to the changing environment to live around humans during the Anthropocene, exploiting the rich and novel resources available among human generated waste.

**Acknowledgements:**

The authors would like to thank Dr Rubina Mondal, Dr Satyaki Mazumder and Dr Udipta Chakraborty for their valuable inputs regarding data analysis. The authors would also like to acknowledge the Indian Institute of Science Education and Research Kolkata for providing infrastructural support. SB would like to thank the University Grants Commission, India for providing him doctoral fellowship.

**References:**

Agudo, R., Rico, C., Vilà, C., Hiraldo, F., & Donzar, J. A. (2010). The role of humans in the diversification of a threatened island raptor. BMC Evolutionary Biology, 10(1). https://doi.org/10.1186/1471-2148-10-384

Allen, M. L., Elbroch, L. M., Wilmers, C. C., & Wittmer, H. U. (2014). Trophic facilitation or limitation? Comparative effects of pumas and black bears on the scavenger community. PLoS ONE, 9(7). https://doi.org/10.1371/journal.pone.0102257

Axelsson, E., Ratnakumar, A., Arendt, M. L., Maqbool, K., Webster, M. T., Perloski, M., Liberg, O., Arnemo, J. M., Hedhammar, Å., & Lindblad-Toh, K. (2013). The genomic signature of dog domestication reveals adaptation to a starch-rich diet. Nature, 495(7441), 360–364. https://doi.org/10.1038/nature11837

Ballejo, F., Lambertucci, S. A., Trejo, A., & de Santis, L. J. M. (2018). Trophic niche overlap among scavengers in Patagonia supports the condor-vulture competition hypothesis. In Bird Conservation International (Vol. 28, Issue 3, pp. 390–402). Cambridge University Press. https://doi.org/10.1017/S0959270917000211

Bascompte, J., & Jordano, P. (2007). Plant-animal mutualistic networks: The architecture of biodiversity. In Annual Review of Ecology, Evolution, and Systematics (Vol. 38, pp. 567–593). https://doi.org/10.1146/annurev.ecolsys.38.091206.095818

Basu, P., Parui, A. K., Chatterjee, S., Dutta, A., Chakraborty, P., Roberts, S., & Smith, B. (2016). Scale dependent drivers of wild bee diversity in tropical heterogeneous agricultural landscapes. Ecology and Evolution, 6(19), 6983–6992. https://doi.org/10.1002/ece3.2360

Bates D, Mächler M, Bolker B, Walker S (2015) Fitting linear mixedeffects models using lme4. J Stat Softw 67:1–48

Beasley, J. C., Olson, Z. H., Selva, N., & DeVault, T. L. (2019). Ecological Functions of Vertebrate Scavenging (pp. 125–157). https://doi.org/10.1007/978-3-030-16501-7_6

Bhadra, A., Bhattacharjee, D., Paul, M., Singh, A., Gade, P. R., Shrestha, P., & Bhadra, A. (2016). The meat of the matter: a rule of thumb for scavenging dogs? Ethology Ecology and Evolution, 28(4), 427–440. https://doi.org/10.1080/03949370.2015.1076526

Bhattacharjee, D., & Bhadra, A. (2020). Humans Dominate the Social Interaction Networks of Urban Free-Ranging Dogs in India. Frontiers in Psychology, 11. https://doi.org/10.3389/fpsyg.2020.02153

Bhattacharjee, D., & Bhadra, A. (2021). Response to short-lived human overcrowding by free-ranging dogs. Behavioral Ecology and Sociobiology, 75(7). https://doi.org/10.1007/s00265-021-03052-x

Cagua, E. F., Wootton, K. L., & Stouffer, D. B. (2019a). Keystoneness, centrality, and the structural controllability of ecological networks. Journal of Ecology, 107(4), 1779–1790. https://doi.org/10.1111/1365-2745.13147

Cagua, E. F., Wootton, K. L., & Stouffer, D. B. (2019b). Keystoneness, centrality, and the structural controllability of ecological networks. Journal of Ecology, 107(4), 1779–1790. https://doi.org/10.1111/1365-2745.13147

Calcagno, V., Mouquet, N., Jarne, P., & David, P. (2006). Coexistence in a metacommunity: The competition-colonization trade-off is not dead. In Ecology Letters (Vol. 9, Issue 8, pp. 897–907). https://doi.org/10.1111/j.1461-0248.2006.00930.x

Campbell, C., Yang, S., Shea, K., & Albert, R. (2012). Topology of plant-pollinator networks that are vulnerable to collapse from species extinction. Physical Review E - Statistical, Nonlinear, and Soft Matter Physics, 86(2). https://doi.org/10.1103/PhysRevE.86.021924

Chakraborty, P., Chatterjee, S., Smith, B. M., & Basu, P. (2021). Seasonal dynamics of plant pollinator networks in agricultural landscapes: how important is connector species identity in the network? Oecologia, 196(3), 825–837. https://doi.org/10.1007/s00442-021-04975-y

Chamberlain, C. P., Waldbauer, J. R., Fox-Dobbs, K., Newsome, S. D., Koch, P. L., Smith, D. R., Church, M. E., Chamberlain, S. D., Sorenson, K. J., & Risebrough, R. (2005). Pleistocene to recent dietary shifts in California condors. www.pnas.orgcgidoi10.1073pnas.0508529102

Christensen RHB (2019). “ordinal—Regression Models for Ordinal Data .” R package version 2019.12-10. https://CRAN.R-project.org/package=ordinal.

Corté S-Avizanda, A., Jovani, R., Carrete, M., Dona´zar, J. A., & Dona´zar, D. (2012). Resource unpredictability promotes species diversity and coexistence in an avian scavenger guild: a field experiment. In Ecology (Vol. 93, Issue 12).

Dormann, C. F. (2011). How to be a specialist? Quantifying specialisation in pollination networks. In Network Biology (Vol. 1, Issue 1). www.iaees.orgArticle

Dormann, C. F. (2022). Using bipartite to describe and plot two-mode networks in R. https://github.com/jgalgarra/bipartgraph,

Estrada, E., 2007. Characterization of topological keystone species local, global and‘‘meso-scale’’ centralities in food webs. Ecol. Complex. 4, 48–57.

Gruber, B., Fründ, J., & Dormann, C. F. (2008). Introducing the bipartite Package: Analysing Ecological Networks Otter Damage Monitoring View project Forces of Extinction in Reptiles View project Introducing the bipartite Package: Analysing Ecological Networks (Vol. 8, Issue 2). https://www.researchgate.net/publication/228861770

Helfield, J. M., & Naiman, R. J. (2001). Effects of salmon-derived nitrogen on riparian forest growth and implications for stream productivity. In Reports Ecology (Vol. 82, Issue 9).

Hewson, R. (1995). Use of salmonid carcasses by vertebrate scavengers. Journal of Zoology, 235, 53–65

Hoornweg, Daniel; Bhada-Tata, Perinaz. 2012. What a Waste : A Global Review of Solid Waste Management. Urban development series;knowledge papers no. 15. World Bank, Washington, DC. © World Bank. https://openknowledge.worldbank.org/handle/10986/17388 License: CC BY 3.0 IGO.

Hoornweg, D., Bhada-Tata, P. & Kennedy, C. Environment: Waste production must peak this century.*Nature* **502,** 615–617 (2013). https://doi.org/10.1038/502615a

Inger, R., Cox, D. T. C., Per, E., Norton, B. A., & Gaston, K. J. (2016). Ecological role of vertebrate scavengers in urban ecosystems in the UK. Ecology and Evolution, 6(19), 7015–7023. https://doi.org/10.1002/ece3.2414

Jagiello, Z., Dylewski, Ł., Tobolka, M., & Aguirre, J. I. (2019). Life in a polluted world: A global review of anthropogenic materials in bird nests. In Environmental Pollution (Vol. 251, pp. 717–722). Elsevier Ltd. https://doi.org/10.1016/j.envpol.2019.05.028

Katlam, G., Prasad, S., Aggarwal, M., & Kumar, R. (2018). Current Science Association Trash on the menu. 115(12), 2322–2326. https://doi.org/10.2307/26978598

Li, C., Li, Q., van Mieghem, P., Stanley, H. E., & Wang, H. (2015). Correlation between centrality metrics and their application to the opinion model. European Physical Journal B, 88(3), 1–13. https://doi.org/10.1140/epjb/e2015-50671-y

Lima, M., Harrington, R., Saldañ, S., Estay, S., Lima, M., Saldaña, S., & Estay, S. (n.d.). Non-linear feedback processes and a latitudinal gradient in the climatic effects determine green spruce aphid outbreaks in the UK. https://doi.org/10.1111/j.2008.0030-1299.16615.x

Luna, Á., Romero-Vidal, P., & Arrondo, E. (2021). Predation and scavenging in the city: A review of spatio-temporal trends in research. In Diversity (Vol. 13, Issue 2, pp. 1–16). MDPI AG. https://doi.org/10.3390/d13020046

Martín González, A. M., Dalsgaard, B., & Olesen, J. M. (2010). Centrality measures and the importance of generalist species in pollination networks. Ecological Complexity, 7(1), 36–43. https://doi.org/10.1016/j.ecocom.2009.03.008

Monroe, B. P., Doty, J. B., Moses, C., Ibata, S., Reynolds, M., & Carroll, D. (2015). Collection and utilization of animal carcasses associated with zoonotic disease in Tshuapa district, the democratic Republic of the Congo, 2012. Journal of Wildlife Diseases, 51(3), 734–738. https://doi.org/10.7589/2014-05-140

Nielsen, A., & Bascompte, J. (2007). Ecological networks, nestedness and sampling effort. Journal of Ecology, 95(5), 1134–1141. https://doi.org/10.1111/j.1365-2745.2007.01271.x

Noreen, Z., & Sultan, K. (2021). Population explosion and behavioural changes of opportunist wild avifauna at a landfill at gujranwala in northeastern punjab: A baseline deviation study. Pakistan Journal of Zoology, 53(6), 2255–2267. https://doi.org/10.17582/journal.pjz/20200211050231

Oksanen J, Kindt R, Legendre P, O'Hara B, Stevens MHH, Oksanen MJ, Suggests MASS (2007) The Vegan package. Commun Ecol Package 10:631–637

Olson, Z. H., Beasley, J. C., Devault, T. L., & Rhodes, O. E. (2012). Scavenger community response to the removal of a dominant scavenger. Oikos, 121(1), 77–84. https://doi.org/10.1111/j.1600-0706.2011.19771.x

Oro, D., Genovart, M., Tavecchia, G., Fowler, M. S., & Martínez-Abraín, A. (2013). Ecological and evolutionary implications of food subsidies from humans. Ecology Letters, 16(12), 1501–1514. https://doi.org/10.1111/ele.12187

Ortiz, N. E., & Smith, G. R. (1994). The production of Clostridium botulinum type A, B and D toxin in rotting carcasses. Epidemiology and Infection, 113(2), 335–343. https://doi.org/10.1017/S0950268800051761

Rooney, N., McCann, K., Gellner, G., & Moore, J. C. (2006). Structural asymmetry and the stability of diverse food webs. Nature, 442(7100), 265–269. https://doi.org/10.1038/nature04887

Sarkar, R., Sau, S., & Bhadra, A. (2019). Scavengers can be choosers: A study on food preference in free-ranging dogs. Applied Animal Behaviour Science, 216, 38–44. https://doi.org/10.1016/j.applanim.2019.04.012

Schlacher, T. A., Weston, M. A., Schoeman, D. S., Olds, A. D., Huijbers, C. M., & Connolly, R. M. (2015). Golden opportunities: A horizon scan to expand sandy beach ecology. Estuarine, Coastal and Shelf Science, 157, 1–6. https://doi.org/10.1016/j.ecss.2015.02.002

Sebastián-González, E., Moleón, M., Gibert, J. P., Botella, F., Mateo-Tomás, P., Olea, P. P., Guimarães, P. R., & Sánchez-Zapata, J. A. (2016). Nested species-rich networks of scavenging vertebrates support high levels of interspecifi c competition. In Ecology (Vol. 97, Issue 1).

Sebastián-González, E., Morales-Reyes, Z., Botella, F., Naves-Alegre, L., Pérez-García, J. M., Mateo-Tomás, P., Olea, P. P., Moleón, M., Barbosa, J. M., Hiraldo, F., Arrondo, E., Donázar, J. A., Cortés-Avizanda, A., Selva, N., Lambertucci, S. A., Bhattacharjee, A., Brewer, A. L., Abernethy, E. F., Turner, K. L., … Sánchez-Zapata, J. A. (2020). Network structure of vertebrate scavenger assemblages at the global scale: drivers and ecosystem functioning implications. Ecography, 43(8), 1143–1155. https://doi.org/10.1111/ecog.05083

Selva, N., & Fortuna, M. A. (2007). The nested structure of a scavenger community. Proceedings of the Royal Society B: Biological Sciences, 274(1613), 1101–1108. https://doi.org/10.1098/rspb.2006.0232

Sen Majumder, S., Chatterjee, A., & Bhadra, A. (2014). A dog's day with humans – time activity budget of free-ranging dogs in India (Vol. 106, Issue 6). https://www.jstor.org/stable/24102275

Sengupta, A., McConkey, K. R., & Kwit, C. (2022). The influence of provisioning on animal-mediated seed dispersal. Oikos, 2022(2). https://doi.org/10.1111/oik.08276

Sharma, N., Gadagkar, R., & Pinter-Wollman, N. (2022). A reproductive heir has a central position in multilayer social networks of paper wasps. Animal Behaviour, 185, 21–36. https://doi.org/10.1016/j.anbehav.2021.12.011

Ulrich, W., Almeida-Neto, M., & Gotelli, N. J. (2009). A consumer's guide to nestedness analysis. In Oikos (Vol. 118, Issue 1, pp. 3–17). https://doi.org/10.1111/j.1600-0706.2008.17053.x

Valente, T. W., Coronges, K., Lakon, C., & Costenbader, E. (n.d.). How Correlated Are Network Centrality Measures?

Wilson, E. E., & Wolkovich, E. M. (2011). Scavenging: How carnivores and carrion structure communities. In Trends in Ecology and Evolution (Vol. 26, Issue 3, pp. 129–135). https://doi.org/10.1016/j.tree.2010.12.011

**Figures with legends:**

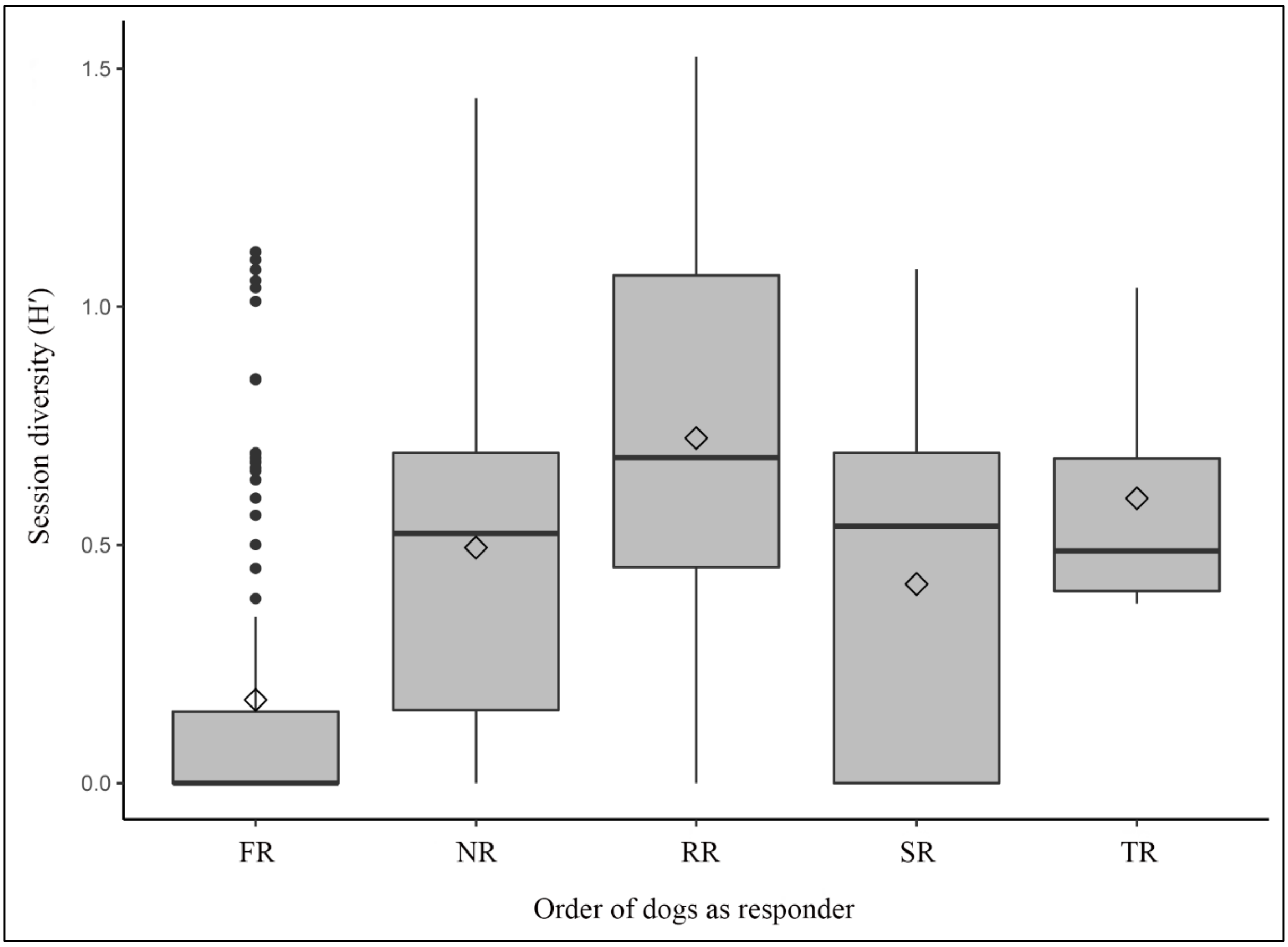

**Figure 1:** A box and whiskers plot showing the effect of dog response order on the session diversity. FR – First responder, SR – Second responder and TR – Third responder, RR – Rest of the positions of Responder; NR – No responder. The black line represents the median value, the open diamond marks the mean, the rectangle shows the 25th and 75th quartile of the data and the whiskers represent the data range.

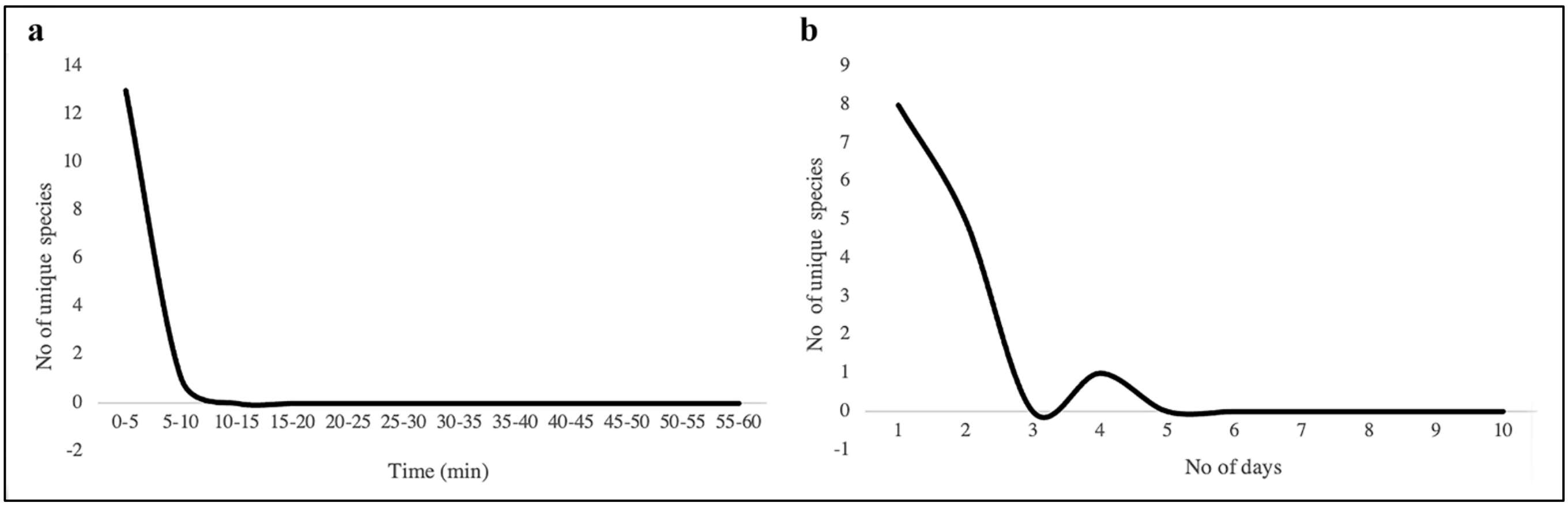

**Figure 2:** Species accumulation curve. a) the number of unique species saturation over time (minutes); b) the number of unique species saturation over the days of the experiment.

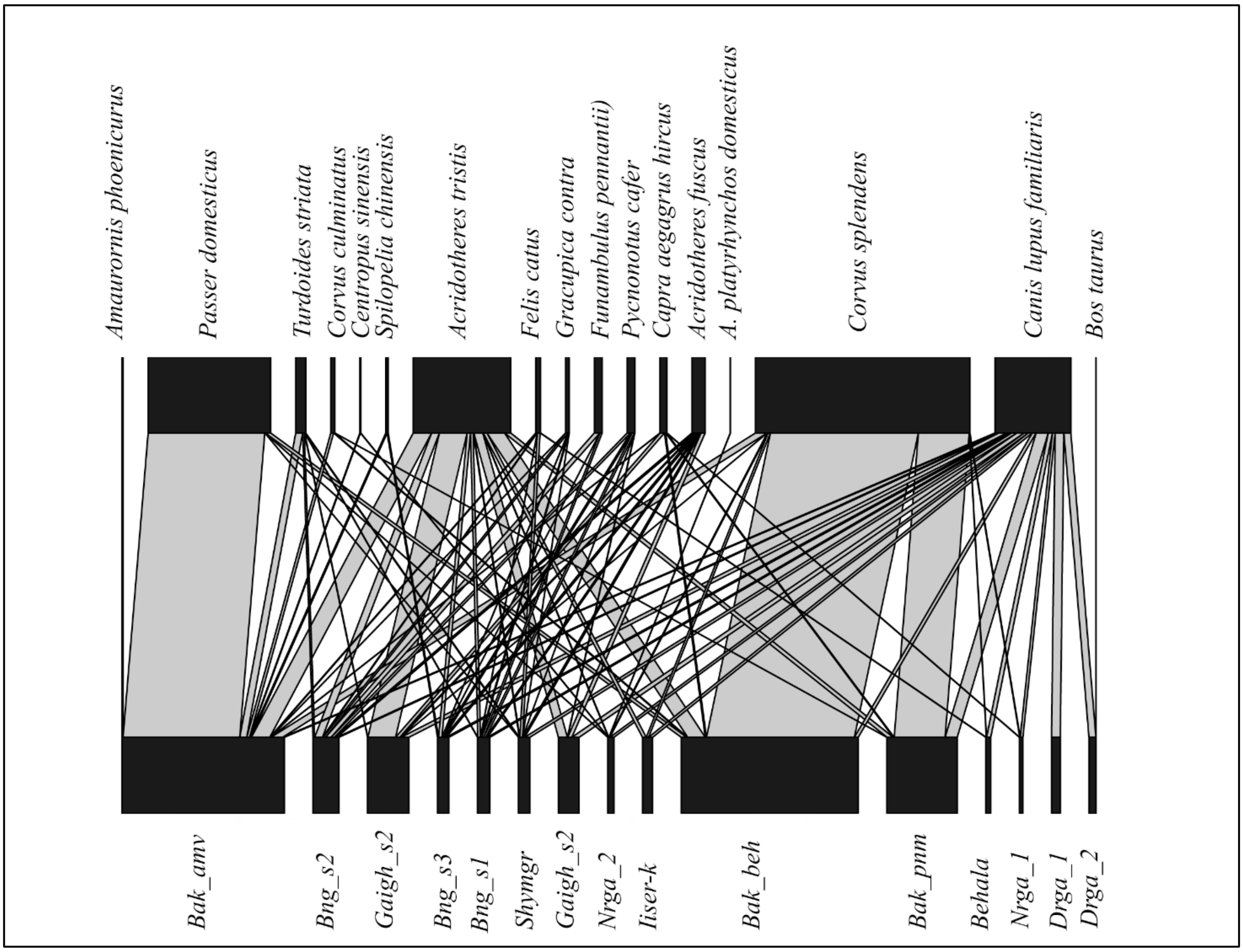

**Figure 3:**

The bipartite graph represents the composite (VF and NVF combined) network structure of scavengers in response to human generated food source in different sites of West Bengal. Each upper rectangle represents a scavenger species and lower rectangle represents a site. The size of the rectangle reflects the no of times a species appears in the network as responder. The lines connect each site with the scavenger species responding at that site, and the thickness of the lines show the number of individuals responding at the respective sites.

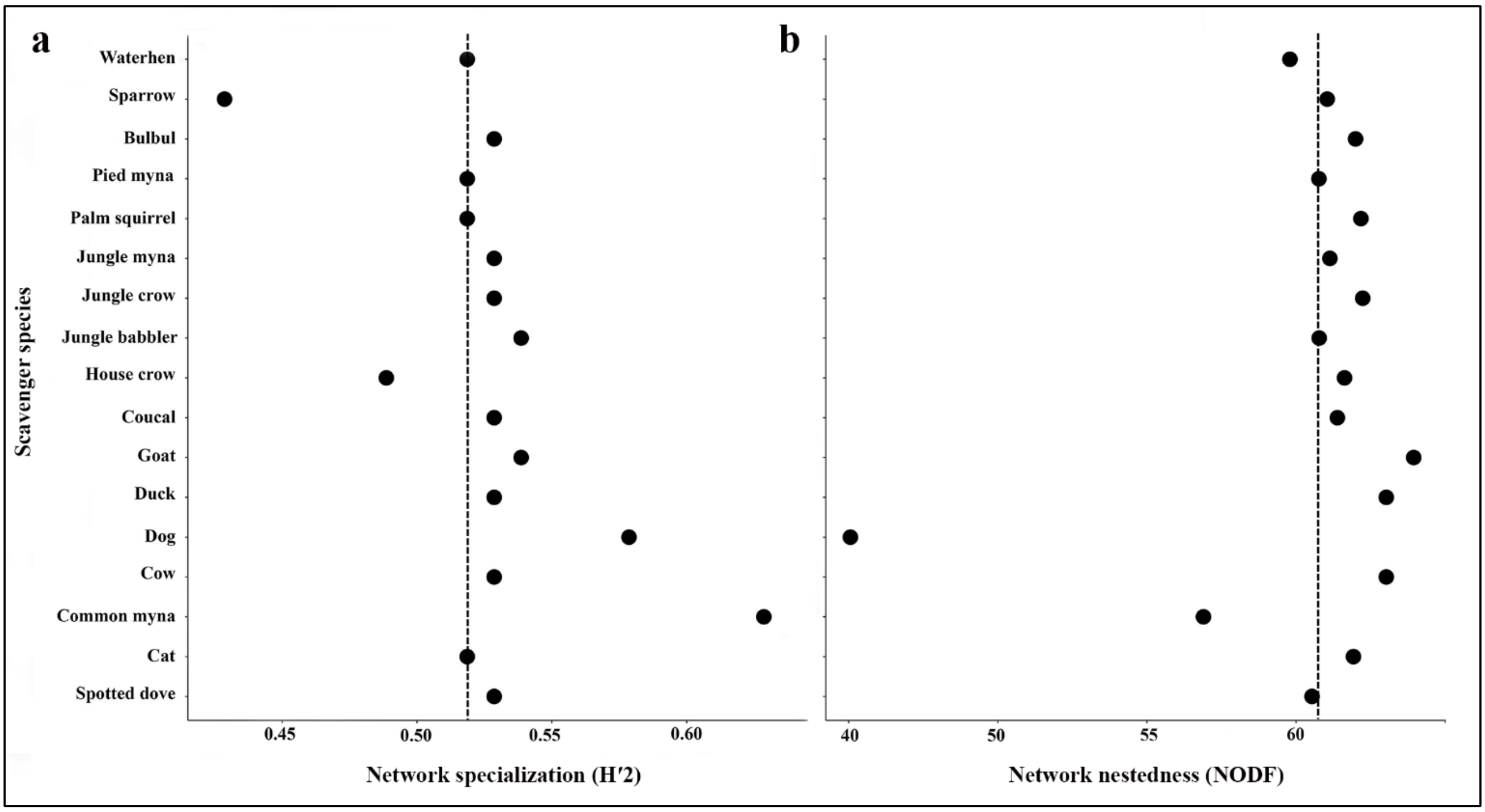

**Figure 4.**

This figure shows the results of virtual exclusion of scavengers describing the contribution of different scavenger species in the network structure: a) network specialization (*H'2* ) and b) network nestedness (NODF) $H'_2$ and NODF values of the respective networks without any exclusion and the dots represent the corresponding values for exclusion of each scavenger species.

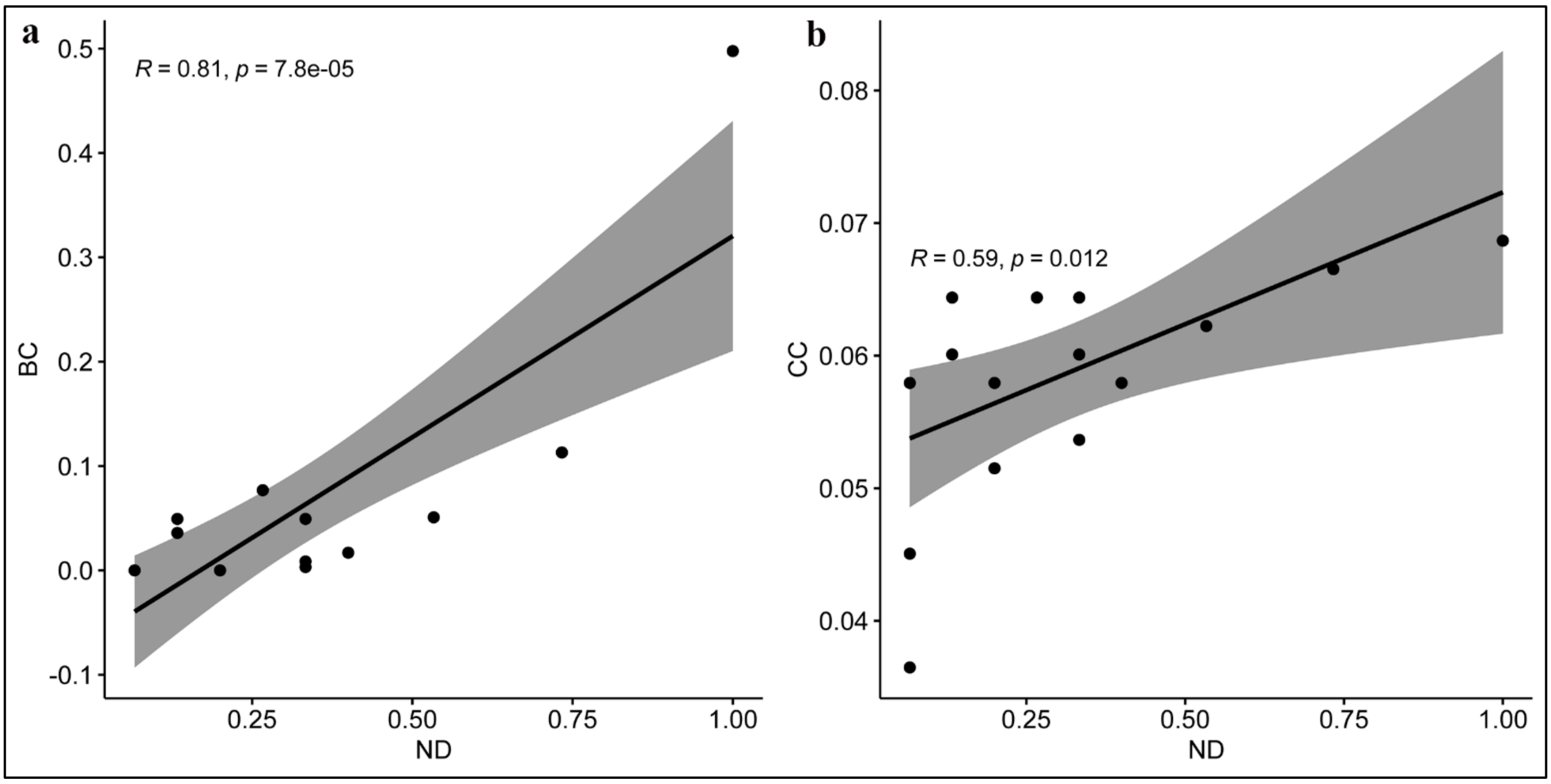

**Figure 5.** a) The relationship between normalized degree (ND) and betweenness centrality (BC) of scavenger species; b) The relationship between normalized degree (ND) and closeness centrality (CC) of scavenger species.

**Supplementary Information**

**Table S1: List of experimental sites**. The table shows the various sites (with GPS locations) in which sampling was carried within the state of West Bengal, India, and the type of habitat (urban/suburban/rural) that they represented.

| **Place** | **Site** | **GPS coordinate** |
|---|---|---|
| Bakkhali | Panchmathani beach | 21.560129, 88.262594 |
| | Bakkhali beach | 21.560129, 88.267003 |
| | Amaravati | 21.560558, 88.266499 |
| Gaighata | Jaleswar | 22.928689, 88.709274 |
| | Srimantapur | 22.931022, 88.722046 |
| Narega | Das para spot 1 | 23.63486 87.96383 |
| | Das para spot 2 | 23.63443, 87.96361 |
| Bangaon | Puraba para | 23.051473, 88.845214 |
| | Nichu Para | 23.052029, 88.842575 |
| | Math para | 23.049635, 88.842064 |
| IISER-Kolkata | ICVS | 22.964909,88.3862325 |
| Behala | Bakultala | 22.485348, 88.301340 |
| Durgapur | Shyampur site 1 | 23.2832046, 87.191320 |
| | Shyampur site 2 | 23.283362, 87.1910723 |
| Shyamnagar | Shibpur | 22.8353262,88.3862325 |

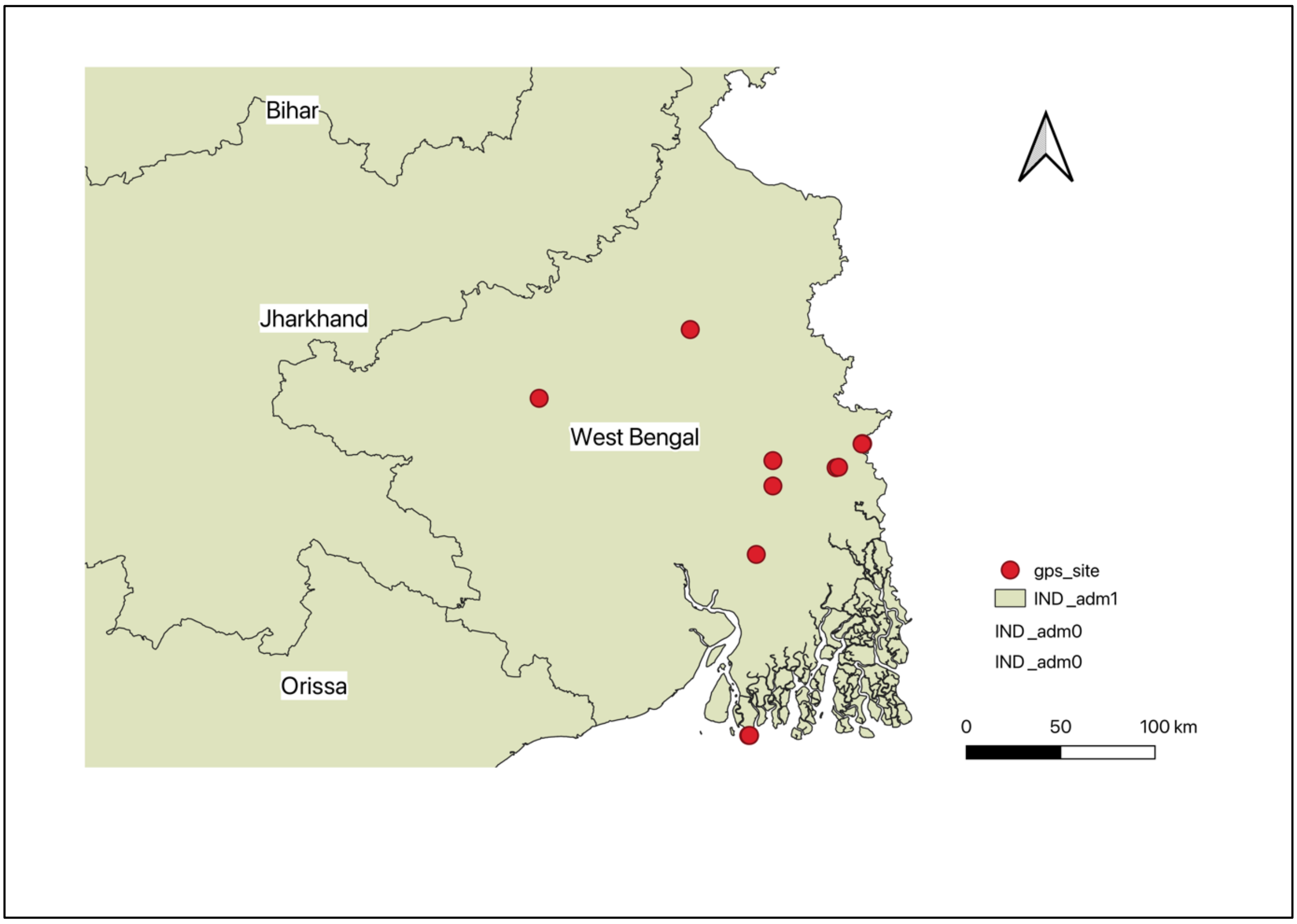

**Figure S1: Map of study sites**. The figure shows the various sites in which sampling was carried within the state of West Bengal, India.

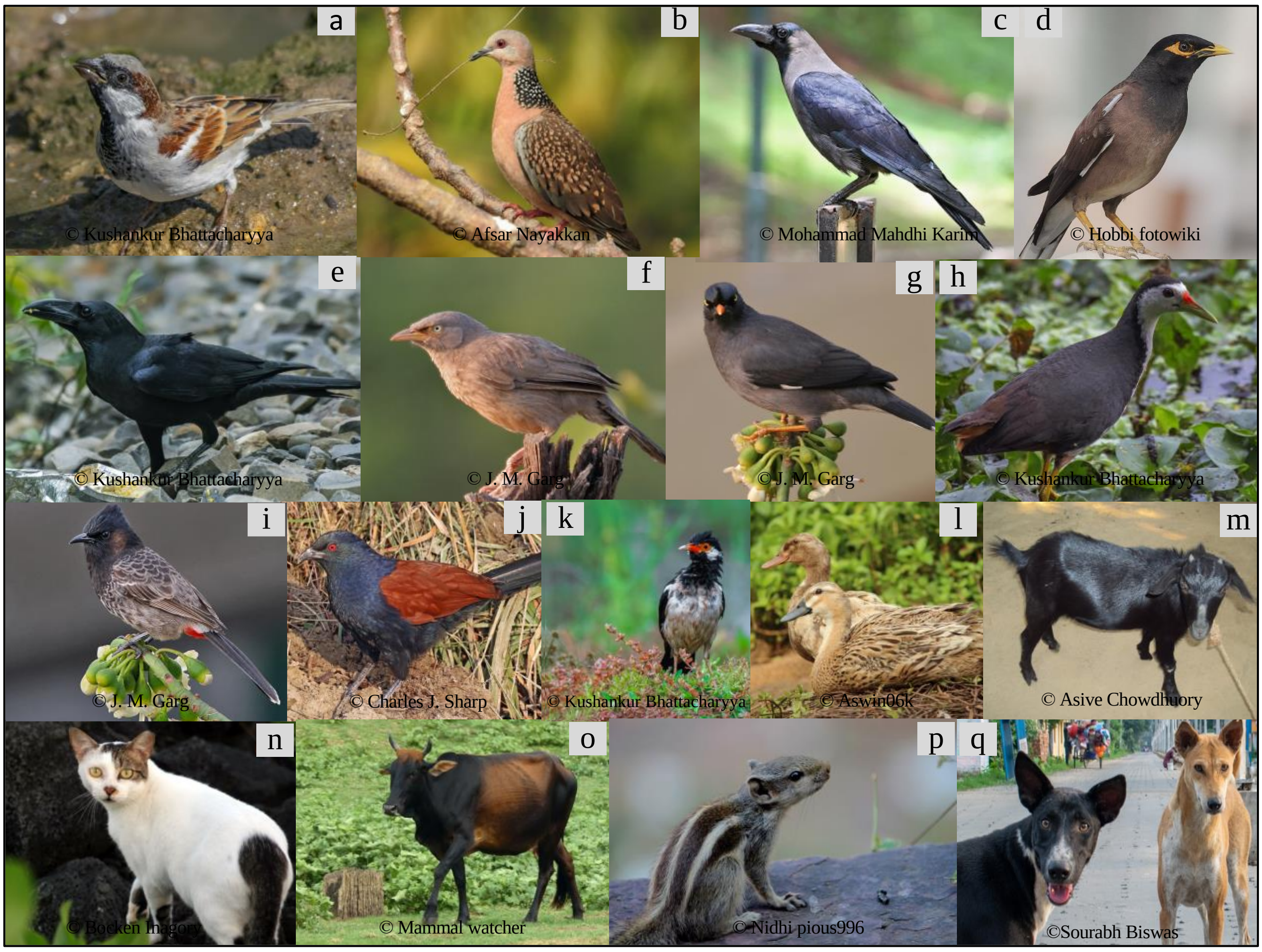

**Figure S2: Scavengers responded in study sites:**

This figure shows the no of scavengers found across different study sites.

a) House sparrow (*Passer domesticus*). b) Spotted dove (*Spilopelia chinensis*). c) House crow (*Corvus splendens*). d) Common myna (*Acridotheres tristis*). e) Jungle crow *(Corvus culminates).* f) Jungle babbler (*Turdoides striatus*). g) Jungle myna (*Acridotheres fuscus*). h) White-breasted waterhen (*Amaurornis phoenicurus*). i) Red-vented bulbul (*Pycnonotus cafer*). j) Greater coucal (*Centropus sinensis*). k) Pied starling (*Gracupica contra*). l) Domestic duck *(Anas platyrhynchos domesticus).* m) Goat (*Capra aegagrus hircus*). n) Cat (*Felis catus*). o) Cow (Bos taurus). p) Palm squirrel (*Funambulus pennantii*).

q) Dog (*Canis lupus familiaris*).

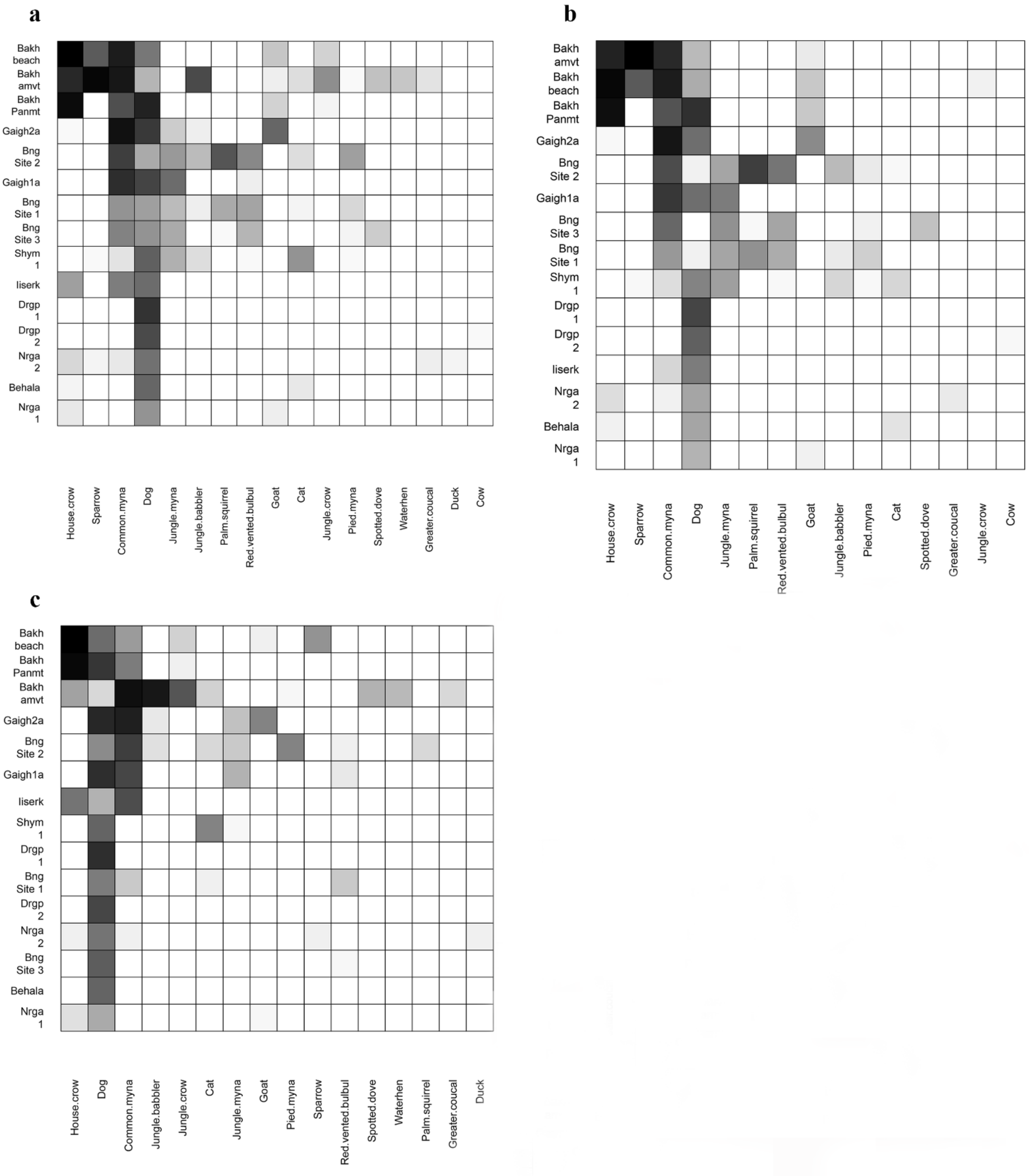

**Figure S3: Response pattern of scavengers in study sites.** This figure shows the no of scavengers found across different study sites. a) Depicts the scavengers found in both VF and VNF combined in the study sites; b) shows the scavengers found only in VF and c) shows the scavengers responded in NVF. The color intensity (black-grey) of the heatmap, figures (a-c) represent the dominance of species response high to low.

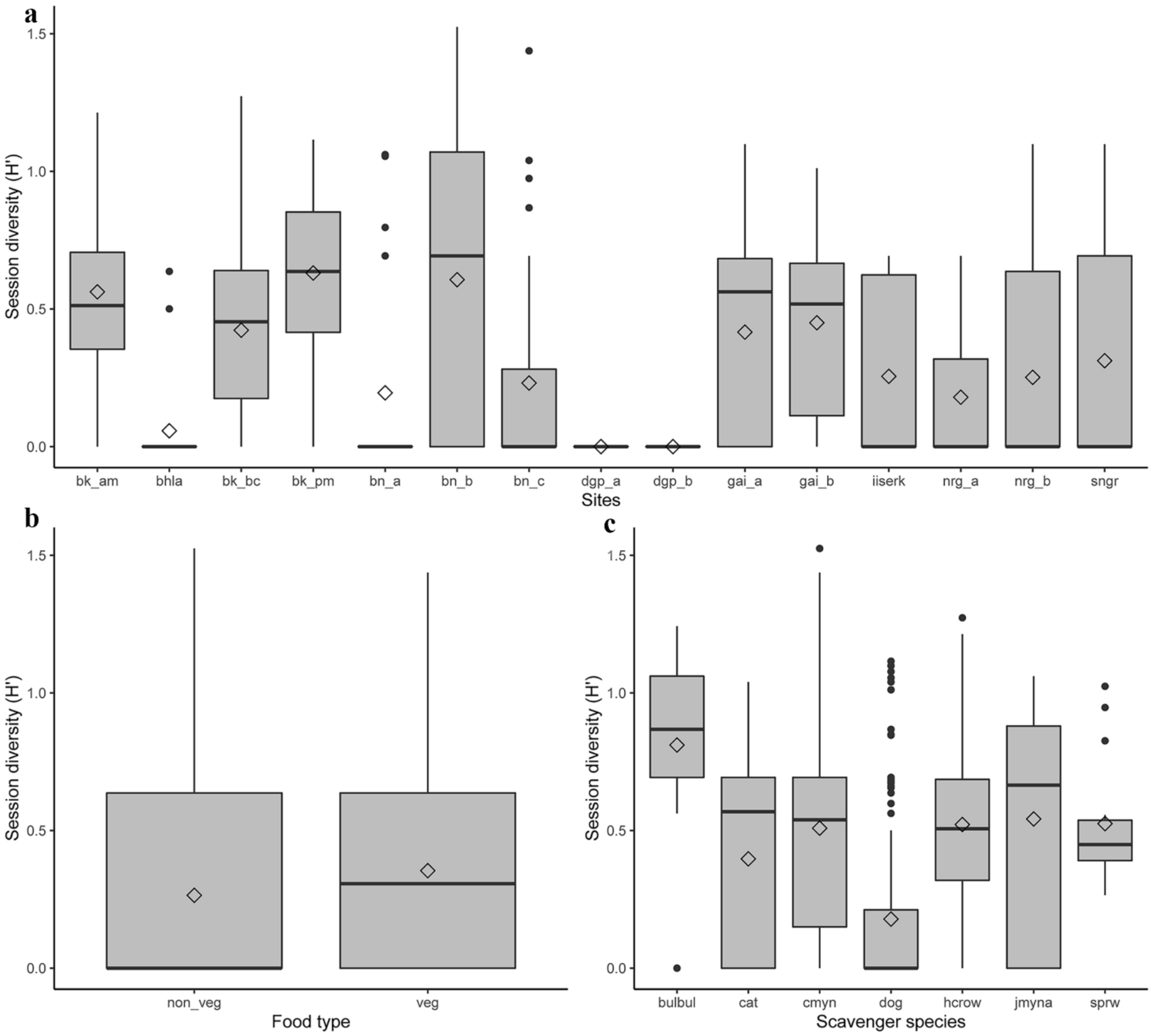

**Figure S4: Difference in session diversity (H′).** a) Describes the difference of session diversity across sites; b) shows the difference of session diversity between veg (VF) and non veg (NVF) food type. Whereas, c) shows the impact of the first responder species on diversity of the session.

**Table S2: Effect of first responder species (scavengers) on the session diversity.**

Ordinal mixed logistic regression (OMLR) analysis.

Model – (Session diversity category (High/Medium and Low) ~ Scavengers as first responder

+ temperature + Session (Morning/ Afternoon), random factor = Study sites)

| Location coefficients: | | | | |
|---|---|---|---|---|
| **Fixed effects** | **Estimate** | **Standard error** | **z value** | **Pr(>\|z\|)** |
| Bulbul | 0.80 | 0.8278 | 0.9644 | 0.3349 |
| cat | 1.41 | 0.5309 | 2.6506 | 0.0080 |
| Common myna | 0.69 | 0.2967 | 2.3311 | 0.0197 |
| House crow | 0.79 | 0.3751 | 2.1111 | 0.0348 |
| Jungle myna | 1.60 | 0.6597 | 2.4203 | 0.0155 |
| sparrow | 0.42 | 0.6561 | 0.6392 | 0.5227 |
| temperature | -0.11 | 0.0494 | -2.2031 | 0.0276 |
| noon (session) | 0.32 | 0.1980 | 1.6002 | 0.1096 |
| Threshold coefficients: | | | | |
| | **Estimate** | **Standard error** | **z value** | |
| High \| Low | -1.9987 | 0.2029 | -9.8496 | |
| Low \| Medium | 1.6206 | 0.1852 | 8.7513 | |
| | | | | |
| **Random effects** | | | | |
| **Group** | **Varience** | **Standard deviation** | | |
| Place | 0.2278142 | 0.4772988 | | |
| | | | | |
| log-likelihood | -402.5226 | | | |
| AIC | 821.0452 | | | |
| Condition number of Hessian | 149.2332 | | | |

**Table S3: Effect of dog response order on session diversity.**

Ordinal mixed logistic regression (OMLR) analysis.

Model – (Session diversity category (High/Medium and Low) ~ Dog response order (First responder/ Second responder/ Third responder/ Rest of the responder and No responder) + Session (Morning/ Afternoon), random factor = Study sites)

| **Location coefficients:** | | | | |
|---|---|---|---|---|
| **Fixed effects** | **Estimate** | **Standard error** | **z value** | **Pr(>\|z\|)** |
| Noon (session) | 0.4427 | 0.1916 | 2.3107 | 0.020851 |
| No Responder (NR) | 1.0825 | 0.2267 | 4.7750 | 1.80E-06 |
| Rest of Responder (RR) | -0.0197 | 0.4951 | 0.0397 | 0.968315 |
| Second Responder (SR) | 1.3127 | 0.6122 | 2.1441 | 0.032024 |
| Third Responder (TR) | -0.4048 | 1.1842 | -0.3419 | 0.732463 |
| **Threshold coefficients:** | | | | |
| | **Estimate** | **Standard error** | **z value** | |
| high\|low | -1.9987 | 0.2029 | -9.8496 | |
| low\|medium | 1.6206 | 0.1852 | 8.7513 | |
| | | | | |
| **Random effects** | | | | |
| **Group** | **Varience** | **Standard deviation** | | |
| place | 0.06402972 | 0.2530409 | | |
| | | | | |
| log-likelihood | -402.5226 | | | |
| AIC | 821.0452 | | | |
| Condition number of Hessian | 149.2332 | | | |

**Table S4: Effect of Common myna response order on session diversity.**

Ordinal mixed logistic regression (OMLR) analysis.

Model – (Session diversity category (High/Medium and Low) ~ Common myna response order (First responder/ Second responder/ Third responder/ Rest of the responder and No responder) + Session (Morning/ Afternoon), random factor = Study sites)

| |
|---|
| **Location coefficients:** |

| **Fixed effects** | **Estimate** | **Standard error** | **z value** | **Pr(>\|z\|)** |
|---|---|---|---|---|
| Noon (session) | 0.4971 | 0.1998 | 2.4875 | 0.0128632 |
| No Responder (NR) | -0.5570 | 0.3616 | -1.5406 | 0.1234131 |
| Rest of Responder (RR) | 0.3089 | 0.4689 | 0.6586 | 0.5101218 |
| Second Responder (SR) | 0.7410 | 0.4900 | 1.5123 | 0.1304526 |
| Third Responder (TR) | 0.5647 | 0.6622 | 0.8526 | 0.3938569 |
| **Threshold coefficients:** | | | | |
| | **Estimate** | **Standard error** | **z value** | |
| high\|low | -2.6135 | 0.3773 | -6.9262 | |
| low\|medium | 1.0006 | 0.3451 | 2.8994 | |
| | | | | |
| **Random effects** | | | | |
| **Group** | **Varience** | **Standard deviation** | | |
| place | 0.3313624 | 0.5756408 | | |
| | | | | |
| log-likelihood | -389.2415 | | | |
| AIC | 794.4831 | | | |
| Condition number of Hessian | 42.75265 | | | |

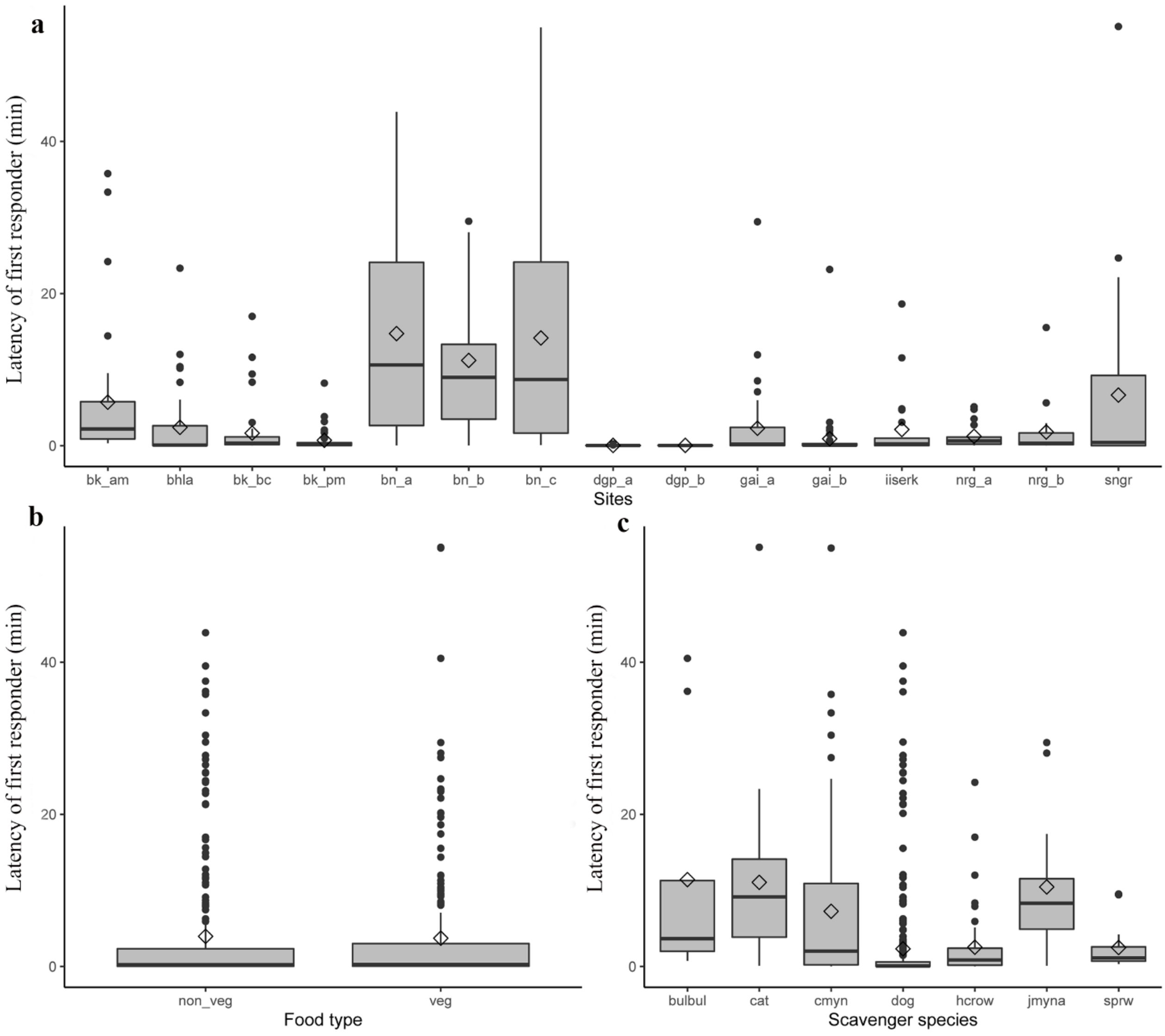

**Figure S5: Difference in latency of first responder (FR).**

Figure a) describes the difference of latency of FR across experimental sites. This Box and Whisker plot also describe the food discovery time of scavenger species in study sites. Figure b) shows the difference of latency of FR between veg (VF) and non veg (NVF) food type. Whereas, c) describes the difference of latency of FR scavengers.

**Table S5: Effect of scavenger species on the food discovery time (Latency of FR)**

Table showing the outcomes of GLMM analysis:

Model – log( Latency of FR) ~ Scavenger species + Temperature + (1| study sites)

| **Fixed effects** | **Estimate** | **Standard error** | **df** | **t value** | **Pr(>\|t\|)** |
|---|---|---|---|---|---|
| (Intercept) | -1.86486 | 0.57747 | 415.84 | -3.229 | 0.00134 ** |
| Bulbul | 0.51253 | 0.26279 | 464.21 | 1.950 | 0.05173 . |
| Cat | 1.03221 | 0.18912 | 464.54 | 5.458 | 7.85e-08 *** |
| Common myna | 0.65557 | 0.10672 | 468.82 | 6.143 | 1.73e-09 *** |
| House crow | 0.54489 | 0.13824 | 471.98 | 3.941 | 9.32e-05 *** |
| Jungle myna | 1.19308 | 0.22937 | 463.94 | 5.202 | 2.97e-07 *** |
| Sparrow | 0.36998 | 0.25370 | 473.00 | 1.458 | 0.14541 |
| Temperature | 0.04138 | 0.01785 | 467.92 | 2.318 | 0.02089 * |
| | | | | | |
| **Random effects** | | | | | |
| Groups | Name | Variance | **Standard deviation** | | |
| place | Intercept | 0.3897 | 0.6243 | | |
| | Residual | 0.5374 | 0.7331 | | |
| AIC | 1142.959 | | | | |
| BIC | 1184.718 | | | | |

*** P < 0.0001, * P < 0.01, . P < 0.05,  P < 0.1

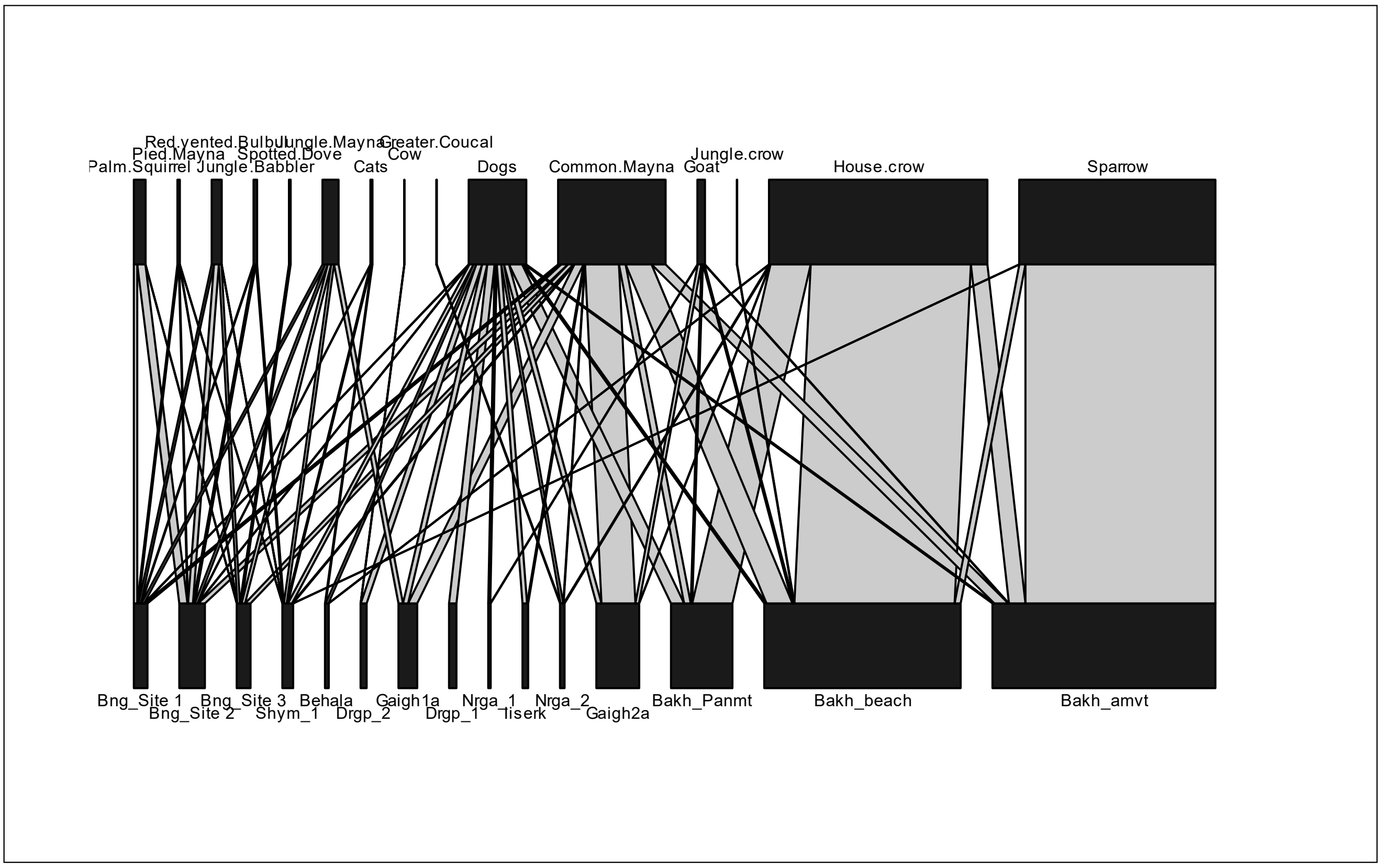

**Figure 6:**

The bipartite graph represents the VF network structure of scavengers in response to human generated food source in different sites of West Bengal. Each upper rectangle represents a scavenger species and lower rectangle represents a site. The size of the rectangle reflects the no of times a species appears in the network as responder. The line match scavenger species responding a specific site, and the width of the line shows the number of individuals responding that specific site.

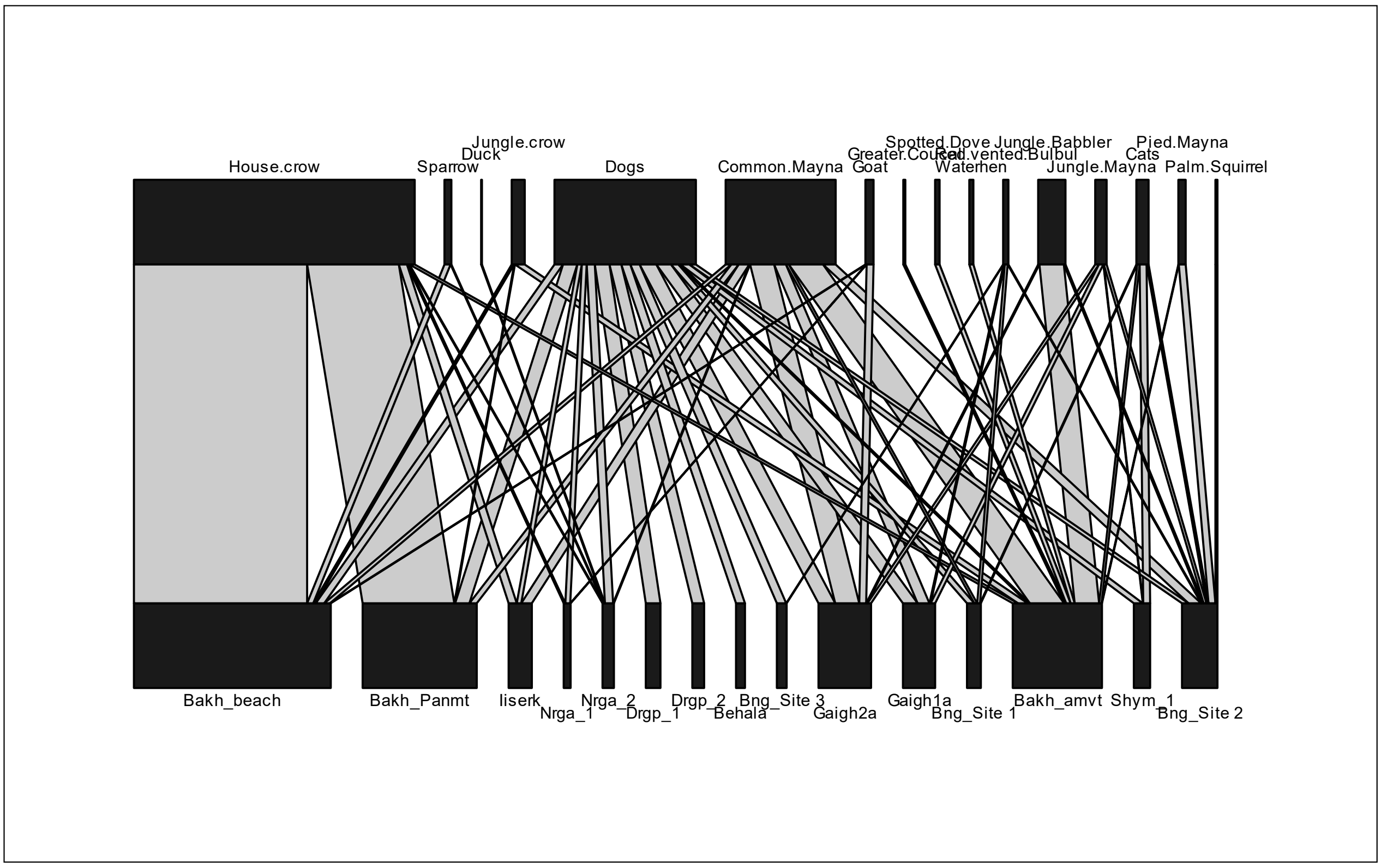

**Figure S7:**

The bipartite graph represents NVF network structure of scavengers in response to human generated food source in different sites of West Bengal. Each upper rectangle represents a scavenger species and lower rectangle represents a site. The size of the rectangle reflects the no of times a species appears in the network as responder. The line match scavenger species responding a specific site, and the width of the line shows the number of individuals responding that specific site.

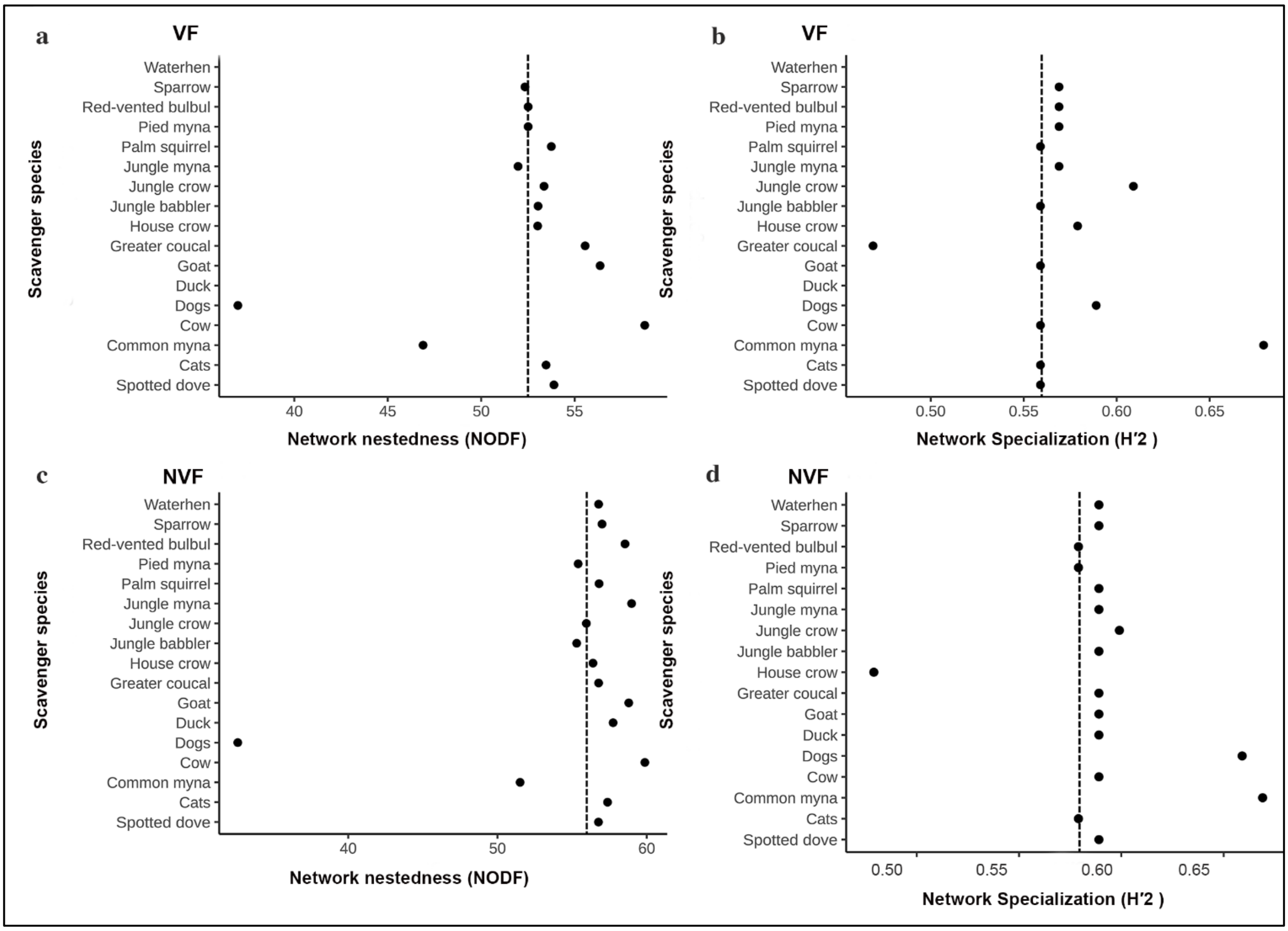

**Figure S8: Virtual exclusion of species.**

This figure shows the results of virtual exclusion scavengers describing the contribution of different scavenger species in the network structure. Fig (a & b) depict the network nestedness (NODF) and network specialization (*H'2* ) in veg food whereas figure (c & d) shows the the network nestedness (NODF) and network specialization (*H'2* ) in non veg food. The dashed line represents the NODF and $H'_2$ value of composite network without any exclusion.

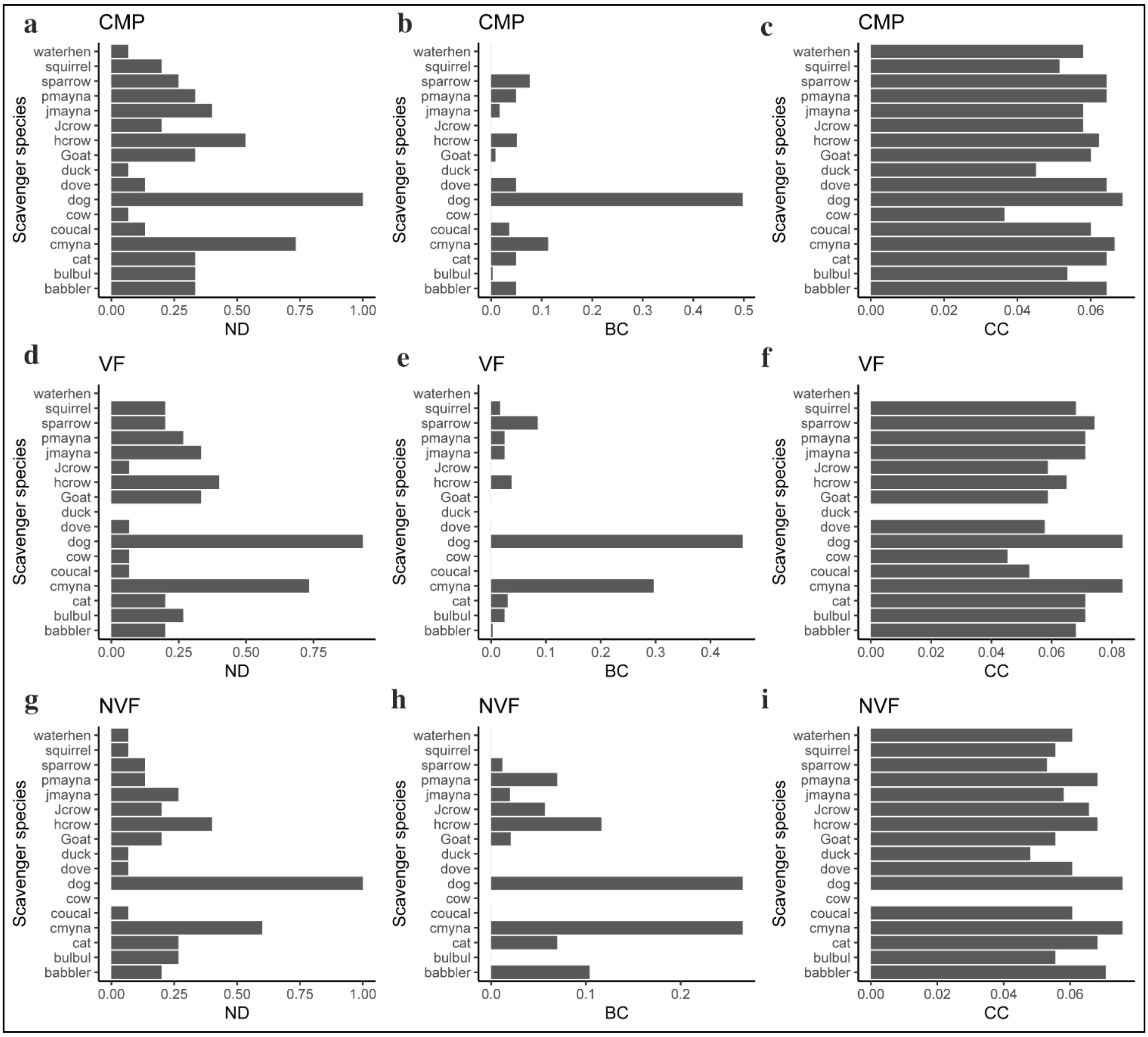

**Figure S9: Species level indices value of scavengers.**

This figure shows the Normality degree (ND), Betweenness centrality (BC) and Closeness centrality (CC) of different scavenger species. Figure (a-c) represent the composite network (CMP), figure (d-f) represent the veg food network (VF) and figure (g-i) represent the non veg food network (NVF) ND, BC and CC accordingly.